\documentclass[journal]{IEEEtran}
\usepackage{times}
\usepackage{subfigure}
\usepackage{latexsym,amsmath,amstext,amsfonts,amssymb,graphicx,epsfig}
\usepackage{algorithm, algorithmic, multirow, xcolor,booktabs}
\usepackage{epstopdf}
\usepackage{cite}
\usepackage{latexsym}
\usepackage{booktabs}
\usepackage{subfigure}

\title{Sparsity-Aware STAP Algorithms Using $L_1$-norm Regularization For Radar Systems}

\author{Zhaocheng~Yang,
Rodrigo~C.~de~Lamare,~\IEEEmembership{Senior Member,~IEEE} and~Xiang Li,~\IEEEmembership{Member~IEEE}
\thanks{Z. Yang and X. Li are with Research Institute of Space Electronics, Electronics Science and Engineering School, National University of Defense Technology, Changsha, 410073, China. e-mail: yangzhaocheng@gmail.com, lixiang01@vip.sina.com.}
\thanks{R. C. de Lamare is with Communications Research Group, Department of Electronics, University of York, YO10 5DD, UK. e-mail: rcdl500@ohm.york.ac.uk}
}

\begin{document}
\maketitle
\begin{abstract}
This article proposes novel sparsity-aware space-time adaptive processing (SA-STAP) algorithms with $l_1$-norm regularization for airborne phased-array radar applications. The proposed SA-STAP algorithms suppose that a number of samples of the full-rank STAP data cube are not meaningful for processing and the optimal full-rank STAP filter weight vector is sparse, or nearly sparse. The core idea of the proposed method is imposing a sparse regularization ($l_1$-norm type) to the minimum variance (MV) STAP cost function. Under some reasonable assumptions, we firstly propose a $l_1$-based sample matrix inversion (SMI) to compute the optimal filter weight vector. However, it is impractical due to its matrix inversion, which requires a high computational cost when in a large phased-array antenna. Then, we devise lower complexity algorithms based on conjugate gradient (CG) techniques. A computational complexity comparison with the existing algorithms and an analysis of the proposed algorithms are conducted. Simulation results with both simulated and the Mountain Top data demonstrate that fast signal-to-interference-plus-noise-ratio (SINR) convergence and good performance of the proposed algorithms are achieved.
\end{abstract}
\section{Introduction}

Space-time adaptive processing (STAP) is an efficient tool for
detection of slow targets by airborne or spaceborne radar systems in
serious environments, such as strong clutter and lots of jammers
\cite{BrennanReed1973,JWard1994,Guerci2003,Melvin2004}. However, the
full-rank adaptive STAP based on linearly constrained minimum
variance (LCMV) criterion gives rise to two of the major limitations
in practical applications of radar \cite{JWard1994,Melvin2004}.
First, the computational load required to solve the interference
matrix inversion is quite intense. In addition, the number of
training data samples required for an accurate estimate of the
interference covariance matrix can become impractical for
high-dimensional problems, particularly in heterogeneous
environments. It is therefore desirable to develop STAP techniques
with low computational complexity and that can provide high
performance in small-sample support situations.

The diagonal loading sample matrix inversion (LSMI) technique is
considered to be a simple and robust approach for both homogeneous
and heterogeneous environments \cite{Blair1988}, but has a high
computational cost. Reduced-rank techniques have been investigated
for solving the previously discussed problems in the last decades
\cite{Pados2001,Pados2007,JScott1997,JScott1999,Scharf2008,Honig2002,Peckham2000,
RuiSTAP2010,Wlei2010, Chang2000,Jiang2010,RuiKA2010,Ruifa2010}. One
of the most important reduced-rank techniques is the class of the
Krylov subspace methods, which includes the auxiliary-vector filters
(AVF) \cite{Pados2001,Pados2007}, the multistage Wiener filter (MWF)
\cite{JScott1997,JScott1999,Scharf2008,Honig2002} and the conjugate
gradient (CG) algorithm
\cite{RuiSTAP2010,Wlei2010,Chang2000,Jiang2010}. These methods
project the observation data onto a lower-dimensional Krylov
subspace and can obtain an improved convergence and tracking
performance. The main differences amongst them lie on the
computational cost, structure of adaptation and ease of
implementation. Knowledge-aided (KA) STAP techniques have currently
gained significant attention as an effective STAP algorithm to
mitigate the effects of the heterogeneity in the secondary data by
exploiting a priori knowledge
\cite{R.Guerci2006,Melvin2006,RuiKA2010,Ruifa2010}. However, the
exact form of prior knowledge is still problem-dependent and hard to
be derived. More recently, several authors have considered
 {sparse recovery (SR)} ideas for moving target
indication (MTI) and STAP problems
 {\cite{SMARIA2006,Jason2010,IvanW2010,KeSun2009,KeSun2010,KeSun2011a,
KeSun2011b}}.  {These work based on SR techniques} relys on the
recovery of the clutter power in the angle-Doppler plane,  {which is
usually carried out via two steps: first, recovering the clutter
angle-Doppler profile by some SR algorithms; second, estimating the
covariance matrix based on the result obtained in the first step,
and computing the Capon's optimal filter. Although some fast sparse
recovery algorithms are proposed, e.g., the fast iterated
shrinkage/thresholding (FISTA) algorithm \cite{IvanW2010}, and the
focal underdetermined system solution (FOCUSS) based algorithm
\cite{KeSun2011a}, it is more computationally expensive than
conventional STAP because of the Capon's optimal filter requiring
matrix inversion, and the recovery procedure being an additive
computational burden.}

{In airborne radar systems, most interference suppression problems
are rank deficient in nature \cite{JWard1994, Guerci2003,
Melvin2004}, that is they require less adaptive degrees of freedom
(DOF) than the full DOF provided by the array. In this case, the
total adaptive DOF provided by the array will be much great than the
number that needed to suppress the interference. Motivated by this,
the authors in \cite{Scott1995} proposed an sequential approach that
gave a sparse solution for the transformation matrix to select the
"best" DOF to be retained in a partially adaptive beamformer.
Moreover, the property described above can be seen that there is a
high degree of sparsity of the filter weight vector. Hence, in our
prior work, an $l_1$ type regularization to the generalized sidelobe
canceler (GSC) STAP processor using the $l_1$-based online
coordinate gradient (OCD) method \cite{ZCYang2011} and the
$l_1$-based recursive least squares method \cite{ZcYangTSP2011} is
introduced to exploit the sparsity of the received data and filter
weights, resulting in an improvement in both convergence rate and
steady-state signal-to-interference-plus-noise ratio (SINR)
performance.} In this paper, we extend the work presented in
\cite{ZCYang2011}  {and \cite{ZcYangTSP2011}} to the direct filter
STAP processor (DFP). By adding the sparsity constraint ($l_1$-norm
regularization) to the MV cost function, we derive the
$l_1$-regularized optimal filter weight vector under some reasonable
assumptions, and then propose a sparsity-aware (SA) adaptive STAP
strategy for airborne radar systems. One direct way is to use the
$l_1$-based SMI recursion algorithm to compute the filter weights.
However, it requires the matrix inversion operation, which prevents
its use in practice. The CG method has a low computational
complexity and is the simplest Krylov subspace method since it only
needs the forward stage, unlike the MWF that requires both forward
and backward stages. Therefore, low complexity $l_1$-based CG type
algorithms are devised. The simulations are conducted using both
simulated and measured data, which show that the proposed algorithms
exhibit improved performance as compared to existing techniques.

This paper is organized as follows. Section II introduces the STAP signal model for airborne radar. In Section III, we first introduce the strategy of the SA-STAP algorithm. Then $l_1$-based SMI and $l_1$-based CG type algorithms are developed and their computational complexity is also shown. Furthermore, we conduct an analysis of the proposed algorithms. In Section IV, some examples of performance of the proposed algorithms with both simulated and the Mountain Top data are exhibited. Finally, the conclusions are given in Section V.

Notation: In this paper, scalar quantities are denoted with italic
typeface. Lowercase boldface quantities denote vectors and uppercase
boldface quantities denote matrices. The operations of
transposition, complex conjugation, and conjugate transposition are
denoted by superscripts $T$, $\ast$, and $H$, respectively. The
symbols $\otimes$ represents the Kronecker product and $\odot$
denotes the Hadamard matrix product. Finally, the symbol
$E\left\{\cdot\right\}$ denotes the expected value of a random
quantity,  {operator $\Re [\cdot]$ selects the real part of
argument, and the symbol $\|\cdot\|_p$ denotes the $l_p$-norm
operation of a vector.}

\section{Signal Model and Problem Statement}
The system under consideration is a pulsed Doppler radar residing on an airborne platform. The radar antenna is a uniformly linear spaced array (ULA) which consists of $M$ elements. The platform is at altitude $h_p$ and moving with constant velocity $v_p$. The chosen coordinate system is shown in Fig.\ref{fig1.sub1}. The angle variables $\phi$ and $ \theta$  refer to elevation and azimuth. The radar transmits a coherent burst of   pulses at a constant pulse repetition frequency (PRF) $f_r=1/T_r$ , where $T_r$ is the pulse repetition interval (PRI). The transmitter carrier frequency is $f_c=c/\lambda_c$, where $c$ is the propagation velocity and $\lambda_c$ is the wavelength. The coherent processing interval (CPI) length is equal to $NT_r$. For each PRI, $K$ time samples are collected to cover the range interval. After matched filtering to the radar returns from each pulse, the received data set for one CPI comprises $KNM$ complex baseband samples, which is referred to as the radar datacube shown in Fig.\ref{fig1.sub2}. The data are then processed at one range of interest, which corresponds to a slice of the CPI datacube. The slice is an $M\times N$ matrix which consists of $M\times1$ spatial snapshots for pulses at the range of interest. It is convenient to stack the matrix column-wise to form the $NM\times1$ vector ${\bf x}[k]$, termed a space-time snapshot, where $k$ is the range sample index and $1\leq k \leq K$ \cite{JWard1994,Guerci2003,Melvin2004}.

Target detection in airborne radar systems can be formulated into a binary hypothesis problem, where the hypothesis $H_0$ corresponds to target absence and the hypothesis $H_1$ corresponds to target presence, given as
\begin{eqnarray}\label{eq1}
\begin{split}
    H_0: {\bf x} = &{\bf x}_u \\
    H_1: {\bf x} = &\alpha_s{\bf s}+{\bf x}_u,
\end{split}
\end{eqnarray}
where $\alpha_s$ is a complex gain and the vector ${\bf s}$, which is the $NM\times1$ normalized space-time steering vector in the space-time look-direction, defined as
\begin{equation}\label{eq2}
    {\bf s}=\frac{{\bf s_t}(f_d)\otimes{\bf s_s}(f_s)}{ {\|{\bf s_t}(f_d)\otimes{\bf s_s}(f_s)\|_2}},
\end{equation}
where ${\bf s_t}(f_d)$ denotes the $N\times1$ temporal steering vector at the target Doppler frequency $f_d$ and ${\bf s_s}(f_s)$ denotes the $M\times1$ spatial steering vector in the direction provided by the target frequency $f_s$. The vector ${\bf x}_u$ encompasses any undesired interference or noise component of the data including clutter ${\bf x}_c$, jamming ${\bf x}_j$ and thermal noise ${\bf x}_n$. Generally, we assume the thermal noise is spatially and temporally uncorrelated, and the jamming is temporally uncorrelated but spatially strongly correlated. As for the clutter, a general model for the clutter space-time snapshot is given by \cite{Melvin2000}
\begin{eqnarray}\label{eq3}
\begin{split}
    {\bf x}_c[k]=&\sum^{N_r}_{m=1}\sum^{N_c}_{n=1}\sigma_{c/k;m,n} \left(\boldsymbol{\alpha_t}(k;m,n) \odot {\bf s_t}(f_{d/k;m,n}) \right)\\
    &\otimes \left(\boldsymbol{\alpha_s}(k;m,n) \odot {\bf s_s}(f_{s/k;m,n})\right),
\end{split}
\end{eqnarray}
where $N_r$ is the number of range ambiguities, $N_c$ is the number of independent clutter patches that are evenly distributed in azimuth about the radar, $\boldsymbol{\alpha_t}(k;m,n)$ is a vector describing the normalized pulse-to-pulse voltages, and $\boldsymbol{\alpha_s}(k;m,n)$ accounts for spatial decorrelation. $\sigma_{c/k;m,n}$ describes the average voltage for the $mn$th clutter patch and $k$th range. The clutter-jammer-noise (for short, calling interference in the following part) covariance matrix ${\bf R}$ can be expressed as
\begin{equation}\label{eq5}
    {\bf R} = E\left\{{\bf x}_u {\bf x}^H_u\right\} = {\bf R}_c + {\bf R}_j +{\bf R}_n,
\end{equation}
where ${\bf R}_c=E\left\{{\bf x}_c {\bf x}^H_c\right\}$, ${\bf R}_j=E\left\{{\bf x}_j {\bf x}^H_j\right\}$ and ${\bf R}_n=E\left\{{\bf x}_n {\bf x}^H_n\right\}$, denote clutter, jammer and thermal noise covariance matrix, respectively.

Generally, the space-time processor linearly combines the elements of the data snapshot, yielding the scalar output\cite{Melvin2004}
\begin{equation}\label{eq6}
    y = {\bf w}^H {\bf x},
\end{equation}
where ${\bf w}$ is the $NM \times1$ weight vector. The idea behind LCMV approach is to minimize the STAP output power whilst constraining the gain in the direction of the desired signal. This leads to the following power minimization with constraints
\begin{eqnarray}\label{eq7}
    \min_{{\bf w}} J({\bf w}) = E\left\{\|{\bf w}^H {\bf x}\|^2_2\right\} \quad    \mbox{s.t.}\quad {\bf w}^H {\bf s} = 1.
\end{eqnarray}
Using the method of Lagrange multipliers, the optimal full-rank LCMV STAP  weights are given by\cite{BrennanReed1973}
\begin{equation}\label{eq8}
    {\bf w}_{\rm LCMV}=\frac{{\bf R}^{-1} {\bf s}}{{\bf s}^H{\bf R}^{-1}{\bf s}}.
\end{equation}

\section{SA-STAP with $L_1$-norm Regularization}

In this section, we detail the design of the proposed SA-STAP strategy, derive the $l_1$-based SMI recursion algorithm and the $l_1$-based CG type algorithms and detail their complexity. Finally, the analysis of the proposed SA-STAP algorithms is shown.

\subsection{SA-STAP Strategy}

 {In airborne radar systems, most interference suppression problems are rank deficient in nature, that is they require less adaptive DOF than are offered by the array, the additional DOF that are not required can be discarded so that only those that are important are retained, which is termed as partially STAP technique \cite{Scott1995}. Furthermore, full DOF will lead to slow convergence, i.e. requiring many snapshots to training the filter, which is difficult to obtain especially in non-homogeneous clutter environments. As a result, the total adaptive DOF provided by the array will be much great than the number that needed to suppress the interference. In another word, there is a high degree of sparsity of the filter weight vector. However, in the practical, it is not easy to estimate the required DOF, related with the sparsity, and to decide which DOF are the most important ones. Herein,} the authors in \cite{ZCYang2011,ZcYangTSP2011} proposed an $l_1$ regularized STAP algorithm for GSC structure to exploit the sparsity of the received data and filter weights. In this paper, we extend this work to a more general framework for airborne radar systems, by employing the sparse regularization to the MV STAP cost function, which is described as the following optimization problem
\begin{equation}\label{eql1.1}
    \min_{{\bf w}} E\left\{\left\|{\bf w}^H{\bf x}\right\|^2_2\right\} + 2\lambda \Gamma\left({\bf w}\right)\qquad \mbox{s.t.}\qquad{\bf w}^H {\bf s} = 1,
\end{equation}
where $\lambda$ is a positive scalar which provides a trade-off between the sparsity and the output interference power. The larger the chosen $\lambda$, the more components are shrunk to zero \cite{Giannakis2010}. The sparse regularization is usually conducted by the $l_0$-norm constraint \cite{MElad2006,MElad2010,Jin2010}. However, since this kind of optimization problem is known to be NP-hard, one of the approximation algorithms, called $l_1$-norm, is considered for the convexity and simple complexity \cite{MElad2010}. In the following, we adopt the $l_1$-norm regularization, i.e., $\Gamma\left({\bf w}\right) = \|{\bf w}\|_1$. Now, the question that arises is how to effectively solve the $l_1$ regularized MV STAP. Albeit convex, the cost function $J_1({\bf w})$ is non-differentiable which leads to difficulty with the use of the method of Lagrange multipliers directly. Thus, we propose an approximation to the regularization term, which is given by
\begin{eqnarray}\label{eql1.3}
    \Gamma\left({\bf w}\right) = \|{\bf w}\|_1 \approx {\bf w}^H\Lambda{\bf w},
\end{eqnarray}
where
\begin{eqnarray}\label{eql1.4}
    {\boldsymbol \Lambda} = \mbox{diag}\left\{\frac{1}{\left|w_1\right|+\epsilon},\frac{1}{\left|w_2\right|+\epsilon},\cdots,\frac{1}{\left|w_{NM}\right|+\epsilon}\right\},
\end{eqnarray}
where $\epsilon$ is a small positive constant  {(e.g.,
$\epsilon=0.01$ is acceptable),and $w_i,i=1,2\cdots,MN$ are the
entries of the filter weight vector ${\bf w}$.}  {Thus the
regularization term ${\bf w}^H\Lambda{\bf w}$ has a quadratic
structure, if we assume the diagonal matrix $\Lambda$ to be fixed.
Minimization can be done iteratively by assuming that the term
$\Lambda$ is fixed, being computed with the current solution ${\bf
w}$ \cite{MElad2006}. So, fixing the term $\Lambda$, we take the
differential term with respect to ${\bf w}^\ast$ of (\ref{eql1.3}),
which is given as follows}
\begin{eqnarray}\label{eql1.5}
    \frac{\partial\|{\bf w}\|_1}{\partial{\bf w}^\ast} \approx \frac{\partial\left({\bf w}^H{\boldsymbol \Lambda}{\bf w}\right)}{\partial{\bf w}^\ast} = \Lambda{\bf w}.
\end{eqnarray}

The above constrained optimization problem described by (\ref{eql1.1}) can be transformed into an unconstrained optimization problem by the method of Lagrange multipliers, whose cost function becomes
\begin{equation}\label{eql1.6}
    \mathcal{L}=E\left\{\left\|{\bf w}^H{\bf x}\right\|^2_2\right\} + 2\lambda \|{\bf w}\|_1 + 2\Re{\left\{\kappa^\ast\left({\bf w}^H{\bf s}-1\right)\right\}},
\end{equation}
where $\kappa$ is a complex Lagrange multiplier. Computing the gradient terms of (\ref{eql1.6}) with respect to ${\bf w}^\ast$ and $\kappa^\ast$, we get
\begin{eqnarray}\label{eql1.7}
\begin{split}
    &\nabla \mathcal{L}_{{\bf w}^\ast} ={\bf R}{\bf w} + \lambda {\boldsymbol \Lambda} {\bf w} + \kappa^\ast {\bf s} \\
    & {\nabla \mathcal{L}_{\kappa^\ast}} = {\bf w}^H{\bf s} - 1.
\end{split}
\end{eqnarray}
By equating the above gradient terms to zero, we obtain the filter weight vector
\begin{equation}\label{eql1.8}
    {\bf w} = \frac{\left[{\bf R}+\lambda{\boldsymbol \Lambda}\right]^{-1}{\bf s}}{{\bf s}^H\left[{\bf R}+\lambda{\boldsymbol \Lambda}\right]^{-1}{\bf s}}.
\end{equation}

By inspecting (\ref{eql1.8}), we verify that there is an additional term $\lambda{\boldsymbol \Lambda}$ in the inverse of the interference covariance matrix ${\bf R}$, which is due to the $l_1$-norm regularization. One should note that the filter weight vector expression in (\ref{eql1.8}) is not a closed-form solution since ${\boldsymbol \Lambda}$ is a function of ${\bf w}$. Thus it is necessary to develop an iterative procedure to compute the filter weight vector, which will be shown in the following parts.

\subsection{$L_1$-Based SMI Recursion Algorithm}
In practice, because the interference covariance is unknown to us, it is most common to compute the interference covariance matrix estimate as \cite{JWard1994,Guerci2003,Melvin2004}
\begin{equation}\label{smi.1}
    \hat{{\bf R}} = \frac{1}{L}\sum^L_{n=1}{\bf x}[n]{\bf x}^H[n],
\end{equation}
where $\left\{{\bf x}[n]\right\}^L_{n=1}$ are known as the secondary or training data. In our following derivation, to develop an iterative procedure, we add an exponential weighting factor to the interference covariance matrix, which may allow the STAP algorithms to accommodate possible non-stationarities in the input. We write the $\hat{{\bf R}}[k]$ as
\begin{equation}\label{smi.2}
    \hat{{\bf R}}[k] = \sum^i_{n=1}\beta^{k-n}{\bf x}[n]{\bf x}^H[n]= \beta\hat{{\bf R}}[k-1] + {\bf x}[i]{\bf x}^H[i],
\end{equation}
where $\beta$ is the forgetting factor, and $\hat{{\bf R}}[0]=\delta{\bf I}$, where $\delta$ is a small positive quantity and ${\bf I}$ is the identity matrix. Since ${\boldsymbol \Lambda}[k]$ is a function of ${\bf w}[k]$, we assume that the filter weight values do not change significantly in a single snapshot step, which is reasonable because we want the instantaneous error of the filter weight vector to change slowly \cite{VanTrees2002}. Hence, ${\boldsymbol \Lambda}[k]$ can be approximated by
\begin{eqnarray}\label{smi.4}
\begin{split}
{\boldsymbol \Lambda}[k] \approx& {\boldsymbol \Lambda}[k-1] = \\ &\mbox{diag}\left\{\frac{1}{\left|w_1[k-1]\right|+\epsilon},\cdots,\frac{1}{\left|w_{NM}[k-1]\right|+\epsilon}\right\}.
\end{split}
\end{eqnarray}
However, we note that the computational complexity of the $l_1$-based SMI recursion algorithm is proportional to $O\left(\left(NM\right)^3\right)$, which is not practical, especially in large size of phased-array antenna. In the next section, we will develop some low complexity algorithms.

\subsection{$L_1$-Based CG  {Algorithms}}
In order to reduce the computational complexity of the $l_1$-based SMI recursion algorithm, we introduce low complexity adaptive algorithms based on CG techniques to iteratively compute the filter weights. There are two different basic strategies for using the CG method. One is the conventional CG (CCG) \cite{Wlei2010,Jiang2010}, which executes several iterations per sample and runs the reset periodically for convergence. The other is the modified CG (MCG) \cite{RuiSTAP2010,Wlei2010,Chang2000}, which operates only one iteration per sample. CCG has a faster convergence than MCG, but a higher computational complexity. In the following, we detail the derivation of the $l_1$-based SA-STAP algorithms based on these two strategies, called $l_1$-based CCG algorithm and $l_1$-based MCG algorithm. For simplicity, we firstly introduce an auxiliary vector given by
\begin{equation}\label{cg.1}
    {\bf v}[k] = \left[\hat{{\bf R}}[k] + \lambda{\boldsymbol \Lambda}[k]\right]^{-1}{\bf r}_t.
\end{equation}
Then the STAP filter weight vector can be described as ${\bf
w}[k]={\bf v}[k]/\left({\bf s}{\bf v}[k]\right)$.  {The solution of
${\bf v}[k]$ described by (\ref{cg.1}) is also the solution of the
following minimal optimization problem \cite{Wlei2010}:}
\begin{equation}\label{cg.2}
    \min \mathcal{J}\left({\bf v}\right) = {\bf v}^H\left[\hat{{\bf R}}+\lambda{\boldsymbol \Lambda}\right]{\bf v} -2\Re\left\{{\bf v}^H{\bf s}\right\}.
\end{equation}
Then the CG-based weight vector is expressed by
\begin{equation}\label{cg.3}
    {\bf v}[k] = {\bf v}[k-1]+\alpha[k]{\bf p}[k],
\end{equation}
where ${\bf p}[k]$ is the direction vector, $\alpha[k]$ is the corresponding adaptive step size.

For the $l_1$-based CCG algorithm, the iteration procedure for the CG-based weight vector ${\bf v}$ is executed per sample. For the $k$th sample, it assumes constant $\hat{{\bf R}}[k] + \lambda{\boldsymbol \Lambda}[k]$ within the internal iterations, and $D$ internal iterations are performed per input data sample. The main difference between the $l_1$-based CCG algorithm and the existing CCG algorithm after the derivation is that we add an additional term $\lambda{\boldsymbol \Lambda}[k]$ to the estimated interference covariance matrix $\hat{{\bf R}}[k]$. A summary of the algorithm is shown in Table \ref{tabel.ccg}.

The $l_1$-based CCG algorithm operates multiple iterations per sample and runs the reset periodically for convergence, which increases the computational load in the sample-by-sample update. In the following, we detail the derivations of the $l_1$-based MCG algorithm with one iteration per sample for STAP. From \cite{Chang2000}, one way to realize the conjugate gradient method with one iteration per snapshot is the application of the degenerated scheme, which means that the residual vector ${\bf g}[k]$ will not be completely orthogonal to the subspace spanned by the direction vectors $\left\{{\bf p}[0],{\bf p}[1],\cdots,{\bf p}[k-1]\right\}$. Under this condition, the adaptive step size $\alpha[k]$ has to fulfill the convergence bound given by
\begin{equation}\label{cg.4}
    0 \leq \left|{\bf p}^H[k]{\bf g}[k]\right|\leq 0.5\left|{\bf p}^H[k]{\bf g}[k-1]\right|,
\end{equation}
where ${\bf g}[k]$ is the negative gradient vector of $\mathcal{J}\left({\bf v}\right)$ in (\ref{cg.2}). Thus, ${\bf g}[k]$ can be written as
\begin{eqnarray}\label{cg.5}
    {\bf g}[k] = -\nabla \mathcal{J}\left({\bf v}\right)_{{\bf v}^\ast} = -\left[\hat{{\bf R}}[k]+\lambda{\boldsymbol \Lambda}[k]\right]{\bf v}[k] + {\bf s},
\end{eqnarray}
which can be calculated recursively by
\begin{eqnarray}\label{cg.6}
\begin{split}
    {\bf g}[k] =& \left(1-\beta\right){\bf s} + \beta{\bf g}[k-1] - \alpha[k]\left[\hat{{\bf R}}[k]+\lambda{\boldsymbol \Lambda}[k]\right]{\bf p}[k]\\
    &-\left\{\lambda\left[1 - \beta\right]{\boldsymbol \Lambda}[k] + {\bf x}[k]{\bf x}^H[k]\right\}{\bf v}[k-1].
\end{split}
\end{eqnarray}
In the previous equation, we use the approximation that ${\boldsymbol \Lambda}[k-1]\approx{\boldsymbol \Lambda}[k]$. Premultiplying (\ref{cg.6}) by ${\bf p}^H[k]$, taking the expectation of both sides and considering ${\bf p}[k]$ uncorrelated with ${\bf s}$, ${\bf x}[k]$ and ${\bf v}[k-1]$ \cite{Chang2000}, we obtain
\begin{eqnarray}\label{cg.7}
\begin{split}
    E\left[{\bf p}^H[k]{\bf g}[k]\right] \approx& \beta E\left[{\bf p}^H[k]{\bf g}[k-1]\right]\\
    &- \beta E\left[{\bf p}^H[k]{\bf s}\right]\\
    &- E\left[\alpha[k]{\bf p}^H[k]\left(\hat{{\bf R}}[k]+\lambda{\boldsymbol \Lambda}[k]\right){\bf p}[k]\right].
\end{split}
\end{eqnarray}
Here, it is assumed that the algorithm converges with the assumption that $E\left[{\bf v}[k-1] - {\bf v}_{opt}\right]\approx0$, $E\left[{\bf x}[k]{\bf r}^H[k]{\bf v}[k-1]\right]\approx{\bf s}$, and $E\left[\lambda\left[1 - \beta\right]{\boldsymbol \Lambda}[k]{\bf v}[k-1]\right]\approx0$. Making a rearrangement of (\ref{cg.7}) and following the convergence bound (\ref{cg.4}), we obtain
\begin{eqnarray}\label{cg.7}
\begin{split}
    \alpha[k] =& \left[{\bf p}^H[k]\left(\hat{{\bf R}}[k]+\lambda{\boldsymbol \Lambda}[k]\right){\bf p}[k]\right]^{-1} \left\{\beta \left[{\bf p}^H[k]{\bf g}[k-1]\right.\right. \\
    & \left.\left.-{\bf p}^H[k]{\bf s}\right] - \mu{\bf p}^H[k]{\bf g}[k-1]\right\}.
\end{split}
\end{eqnarray}
where $0\leq\mu\leq0.5$. The direction vector is a linear combination from the previous direction vector and the negative gradient, which is described as
\begin{equation}\label{cg.8}
    {\bf p}[k] = {\bf g}[k-1] + \nu[k]{\bf p}[k],
\end{equation}
where $\nu[k]$ is computed for avoiding the reset procedure by employing the Polak-Ribiere approach, which should have an improved performance \cite{Chang2000,Wlei2010}, and is stated as
\begin{equation}\label{cg.9}
    \nu[k] = \frac{\left[{\bf g}[k]-{\bf g}[k-1]\right]^H{\bf g}[k]}{{\bf g}^H[k-1]{\bf g}[k-1]}.
\end{equation}
The proposed $l_1$-based MCG STAP algorithm is summarized in Table \ref{tabel.mcg}.

 {From above discussions, two aspects should be noted that: First, the performance of our proposed algorithms (both $l_1$-based SMI and $l_1$-based CG-type algorithms) depends on regularization parameter $\lambda$. An approach to choose $\lambda$ is introduced in \cite{ZcYangTSP2011}, which can be easy to extend to our proposed algorithms, but not discussed in this paper for saving space. Second, the convergence analysis in \cite{Wlei2010} is suitable to our proposed CG-type algorithms, where the convergence is governed by
\begin{equation}\label{cg.10}
    \|{\boldsymbol \varsigma}_{i+1}[k]\|_{{\bf G}[k]} \leq 2\left(\frac{\sqrt{\tau_{\rm max}/\tau_{\rm min}}-1}{\tau_{\rm max}/\tau_{\rm min}+1}\right)^i\|{\boldsymbol \varsigma}_{0}[k]\|_{{\bf G}[k]},
\end{equation}
where ${\boldsymbol \varsigma}_{i}[k]={\bf v}_{\rm opt}[k] - {\bf v}_{i}[k]$ is the CG-based weight vector error at the $i$th iteration for the $k$th snapshot, ${\bf v}_{\rm opt}[k]$ is the optimal solution at the $k$th snapshot, $\tau_{\rm max}$ and $\tau_{\rm min}$ are the maximal and minimal eigenvalues with respect to ${\bf G}[k]=\hat{\bf R}[k] + \lambda{\boldsymbol \Lambda}[k]$, and $\|{\boldsymbol \varsigma}_{i}[k]\|_{{\bf G}[k]}={\boldsymbol \varsigma}^H_{i}[k]{\bf G}[k]{\boldsymbol \varsigma}_{i}[k]$. From the above equation, we note that the convergence behavior of the proposed algorithms is related to the CG-based weight vector error ${\boldsymbol \varsigma}_{0}[k]$ and the condition number $\tau_{\rm max}/\tau_{\rm min}$.
}

\subsection{Complexity Analysis}
\begin{figure}[!htb]
\centering
\includegraphics[width=86mm]{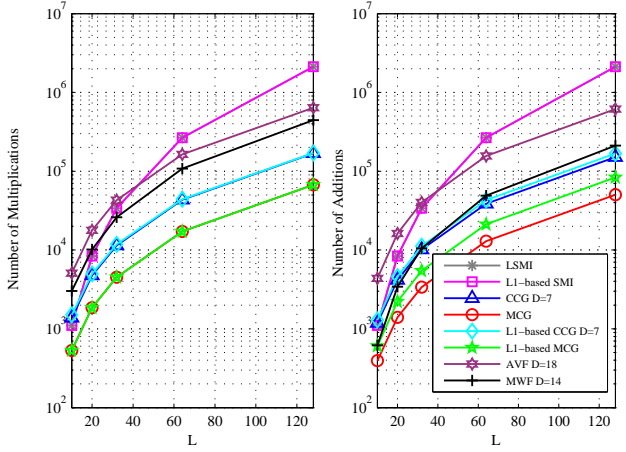}
\caption{
The computational complexity per snapshot.} \label{complexity}
\end{figure}

In this section, we detail the computational complexity in terms of
complex additions and complex multiplications of the proposed
$l_1$-based SMI, $l_1$-based CG type algorithms, and other existing
STAP algorithms, namely the LSMI, the AVF, the MWF and the
conventional CG type algorithms, as shown in Table
\ref{table.complexity}.  {One aspect should be noted that, the rank
$D$ may not equal to the clutter rank, and can be smaller than that.
This is because the principle of the Krylov subspace approach is
different from that of the eigen-decomposition approach. An
eigen-decomposition approach would usually require an SVD on the
full-rank covariance matrix and the selection of the $D$
eigenvectors associated with the $D$ largest eigenvalues, which is
high related to the clutter rank. In contrast to that, the
Krylov-based approach does not require eigen-decomposition and
selects the $D$ basis vectors which minimize the desired cost
function and will form the projection matrix, where $D$ can be
decreased without significantly degrading the SINR
\cite{Honig2002}.} In the table, $D$ is the rank for CCG type, AVF
and MWF algorithms, and $L=NM$ is the system size. Seen from the
table, the computational complexity of $l_1$-based SMI is similar to
the conventional LSMI algorithm, both requiring one to calculate the
matrix inversion. With respect to the proposed $l_1$-based CG type
algorithms, the computational complexity is nearly the same as the
conventional CG type algorithms. Note that the complexity of CCG
type, AVF and MWF algorithms is dependent on the rank $D$. This is a
tradeoff between complexity and performance. We found that the rank
of the proposed $l_1$-based CCG algorithm with $D=7$ works well
(while the best rank for AVF and MWF is much larger), as will be
verified in the following simulations. The low-rank characteristic
will bring computational savings. The computational complexity of
all algorithms is shown in Fig.\ref{complexity}, where we use the
best rank obtained from the simulations for these algorithms ($D=7$
for CCG type, $D=18$ for AVF and $D=14$ for MWF). We see that the
proposed CG type algorithms have much lower complexity than AVF and
MWF algorithms.

Furthermore, it requires to compute the filter weights repeatedly for target detection in airborne radar systems, especially in heterogeneous environment. In this case, our proposed algorithms can work in an iterative way and do not need to recompute all the filter weights, which can lead to significant computational savings. Usually, secondary data of the sliding window are used in detection procedures, where the parameter that defines the length of the sliding window is $K$. Assume ${\bf R}_i[K]$ denotes the estimated interference covariance matrix according to (\ref{smi.2}) and ${\bf w}_i[K]$ denotes the filter weight vector at the cell under test (CUT) of the $i$th range bin, respectively. Consider the case of the $i+1$th CUT, we first remove the impact of $i+1$th CUT, given by
\begin{equation}\label{com.1}
    \beta{\bf R}_{i+1}[K-1] = {\bf R}_i[K]-{\bf x}[i+1]{\bf x}^H[i+1].
\end{equation}
Since an exponentially decaying data window is used, we do not need to remove the first snapshot used to compute the filter weights. Then, similarly, we consider the case of adding snapshots. Two snapshots, one is at the primary $i$th CUT and another is the new snapshot ${\bf x}_{\rm new}$ which was not included in the sliding window before, should be added to the $i+1$th CUT secondary data. The procedure can be written as
\begin{equation}\label{com.2}
\begin{split}
    {\bf R}_{i+1}[K] = &\beta{\bf R}_{i+1}[K-1]+\\
    &\beta{\bf x}_{\rm new}{\bf x}^H_{\rm new} + {\bf x}[i+1]{\bf x}^H[i+1].
\end{split}
\end{equation}
As for the filter weight vector ${\bf w}_{i+1}[K]$ at the $i+1$th CUT, it can be updated using the new interference covariance matrix and the filter weight vector ${\bf w}_{i}[K-1]$.

In addition, the proposed algorithms adopt an adaptive filtering approaches, which can obtain a near optimum interference rejection at a low cost \cite{Klemm1998}. The advantage of this approach is that filtering can be accomplished in a pipeline mode as the echo pulses come in. The required number of calculations for filtering can be realized easily with nowadays digital technology \cite{Diniz2002}.

\subsection{Analysis of the SA-STAP Algorithm}
At this point, we have finished the derivation of the SA-STAP algorithms. The following simulation results will show that the proposed SA-STAP algorithms have a faster SINR convergence speed and better SINR steady-state performance than the conventional algorithms. This translates into a superior detection performance. However, why do the SA-STAP algorithms work is an interesting question. This section will try to explain that from two points of view.

First, to understand the behavior, we write the filter weight vector using the eigenvalue decomposition (EVD) of $\hat{{\bf R}}$. We assume that the eigenvalues of the estimated interference covariance matrix are $\hat{\gamma}_n$ with the corresponding eigenvectors denoted by ${\bf u}_n$, $n=1,2,\cdots,NM$. The eigenvalues are ordered as,
\begin{equation}\label{a.1}
    \hat{\gamma}_1\geq\hat{\gamma}_2\geq\cdots\geq\hat{\gamma}_{NM}=\hat{\gamma}_{min}.
\end{equation}
Thus, through the EVD, the estimated interference covariance matrix can be written as
\begin{eqnarray}\label{a.r}
    \hat{{\bf R}} = \sum^{NM}_{n=1}\hat{\gamma}_n{\bf u}_n{\bf u}^H_n.
\end{eqnarray}
Substituting (\ref{a.r}) into (\ref{eql1.8}), the filter weight vector of the SA-STAP algorithm can be written as
\begin{eqnarray}\label{a.4}
    {\bf w}_{SA} = \varsigma_{SA}\left\{{\bf r}_t -
    \sum^{NM}_{n=1}\frac{\hat{\gamma}_n+\Delta_n-\hat{\gamma}_{min}}{\hat{\gamma}_n +\delta^{SA}_{min}+\Delta_n} \left({\bf u}^H_n{\bf r}_t\right) {\bf u}_n\right\},
\end{eqnarray}
where $\delta^{SA}_{min} = \min\left(\frac{\lambda}{\left|w_n\right|+\epsilon}\right), n=1,2,\cdots,NM$, $\Delta_n$ is the difference between $\frac{\lambda}{\left|w_n\right|+\epsilon}$ and $\delta^{SA}_{min}$, and $\varsigma_{SA}$ is a scalar quantity, which does not affect the SINR.

By inspecting (\ref{a.4}), we observe that the SA-STAP belongs to the class of diagonal loading STAP techniques in a sense. Moreover, it is equivalent to an adaptive diagonal loading technique, which will apply to each eigenbeam of the interference covariance matrix different weights and exploit the sparsity of the filter weights and the received data.

\begin{figure}[!htb]
\centering
\includegraphics[width=86mm]{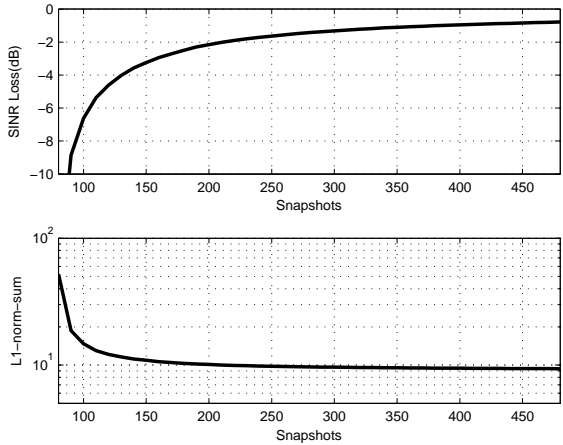}
\caption{
The relationship between the SINR performance and the $l_1$-norm-sum quantity of the filter weights.} \label{sinr_weight}
\end{figure}

Second, we will investigate the relationship between the SINR performance and the $l_1$-norm-sum quantity of the filter weights. Assume the scene is the same as the one with homogeneous environment introduced in the next section. We compute the SINR loss and the $l_1$-norm-sum quantity of the filter weights against the number of snapshots using the SMI algorithm. The results are plotted in Fig.\ref{sinr_weight}. From the figure, we find that the better the SINR performance, the smaller the $l_1$-norm-sum quantity of the filter weights. From this point of view, a constraint on the $l_1$-norm-sum quantity of the filter weights can lead to fast convergence, which in fact exploits the sparsity of the received data and filter weights.

\section{Performance Assessment}
In this section, we assess the proposed SA-STAP algorithms using both simulated and measured data and compare them with the existing algorithms, such as the conventional CCG, MCG, MWF, AVF and LSMI algorithms. We measure the SINR, the SINR loss and the probability of detection curves, where the SINR and the SINR loss are defined as follows \cite{Melvin2004}, respectively.
\begin{eqnarray}\label{eqsinr}
    \mbox{SINR} = \frac{\left|\hat{\bf w}^H{\bf x}\right|^2}{\left|\hat{\bf w}^H{\bf R}\hat{\bf w}\right|},
\end{eqnarray}
\begin{eqnarray}\label{eqsinrloss}
    \mbox{SINR}_{\rm loss} = \frac{\left|\hat{\bf w}^H{\bf x}\right|^2}{\left|\hat{\bf w}^H{\bf R}\hat{\bf w}\right|\left|{\bf s}^H{\bf R}^{-1}{\bf s}\right|},
\end{eqnarray}
where ${\bf R}$ is the exact interference covariance matrix at the detection range bin and $\hat{\bf w}$ is the estimated filter weights using the neighbor secondary data.
\subsection{Simulated Data}
Consider a monostatic sidelooking radar with $M=10$ antenna elements
and $N=8$ pulses in one CPI, giving a space-time steering vector of
length $L=80$. We assume a simulated scenario with the following
parameters: half-wavelength spaced antennas, uniform transmit
pattern, carrier frequency $450$MHz, PRF set to $300$Hz, platform
velocity of $50$m/s and height of $9000$m, the clutter uniformly
distributed from azimuth $-\pi/2$ to $\pi/2$ with
clutter-to-noise-ratio (CNR) of $40$dB, two jammer located at $-45$
and $60$ with jammer-to-noise-ratio (JNR) of $40$dB, the target
located at $0^\circ$ azimuth with Doppler frequency of $100$Hz and
signal-to-noise-ratio (SNR) of $0$dB, and the thermal noise power is
$0.01$W. We consider the inner clutter motion (ICM) in simulated
data. One common model, referred to as the Billingsley model, was
developed by Billingsley of MIT Lincoln Laboratory
\cite{Guerci2003}.  {The only parameters required to specify the
clutter Doppler power spectrum are essentially the shape parameter
$b$ and the wind speed parameter $\omega$. In this paper, we assume
$b=3.8$ and $\omega=51.45$ miles per hour (mph).} All presented
results are averages over 100 independent Monte Carlo runs.

\begin{figure}[!htb]
\centering
\includegraphics[width=86mm]{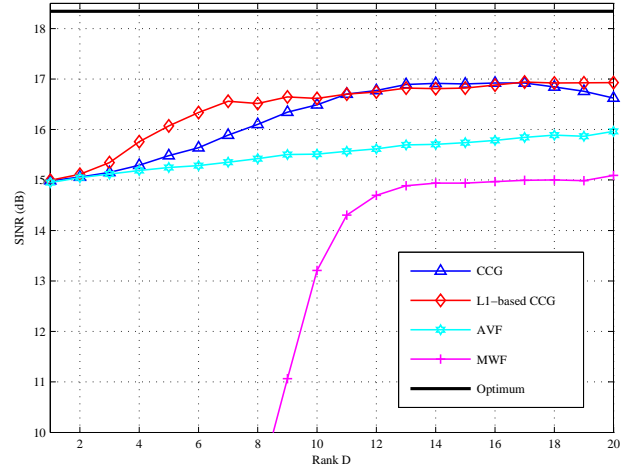}
\caption{
The SINR performance versus the rank $D$. Parameters: the diagonal loading factor for AVF and MWF algorithms is $10$dB to the thermal noise power; $\beta=0.9998$, $\eta_{\rm CCG}=10^{-5}$ and ${\bf R}[0]=0.001{\bf I}$ for CCG type algorithms; $\lambda=2$ for $l_1$-based CCG algorithms.} \label{rank_homo}
\end{figure}

In our first example, we consider the SINR performance versus the rank $D$ of the proposed $l_1$-based CCG algorithm, the conventional CCG algorithm, the AVF algorithm and the MWF algorithm. A total of $K=160$ snapshots are considered. The results in Fig.\ref{rank_homo} show that our proposed $l_1$-based CCG algorithm can obtain its best performance when the rank is larger than $D=7$. It is much lower rank to obtain its best performance than that of AVF ($D=18$) and MWF ($D=14$) algorithms. The low-rank characteristic will bring considerable computational savings, which is very important for STAP in radar systems. One should note that the performance of the conventional CCG algorithm will degrade when the rank is too large, while our proposed $l_1$-based CCG can always keep good performance resulting in further robustness. Since the SINR performance is much worse when the rank is lower than the best rank, thus, we will use $D=7$ for CCG type algorithms, $D=18$ for the AVF algorithm and $D=14$ for the MWF algorithm in the following examples.

\begin{figure}[!htb]
\centering
\includegraphics[width=86mm]{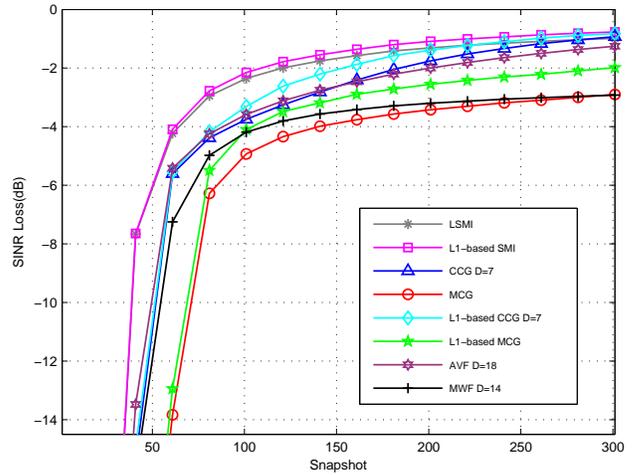}
\caption{
The SINR loss performance against the number of snapshots $K=320$. Parameters: the diagonal loading factor for LSMI, AVF and MWF algorithms is $10$dB to the thermal noise power; $\beta=0.9998$ and ${\bf R}[0]=0.001{\bf I}$ for CG type algorithms; $\lambda=1$ for the $l_1$-based SMI, $\lambda=2$ for the $l_1$-based CCG algorithm and $\lambda=1$ for the $l_1$-based MCG algorithm; $\eta_{\rm CCG}=10^{-5}$.} \label{sinr_snapshots_homo}
\end{figure}

In the next example, we evaluate the SINR loss performance against the number of snapshots $K=320$ of the proposed algorithms with the existing algorithms, as depicted in Fig.\ref{sinr_snapshots_homo}. The curves show that: (1) the SINR performance of the proposed $l_1$-based SMI algorithm is a suboptimal algorithm, but exhibits the best performance compared with other algorithms. (2) $l_1$-based CG type algorithms outperform conventional CG type algorithms in terms of convergence rate and steady-state performance; (3) the SINR performance of the $l_1$-based CCG algorithm is better than AVF and MWF algorithms. (4) Although the $l_1$-based MCG algorithm shows slower SINR convergence than the MWF algorithm, we can obtain a better SINR performance when the number of snapshots is larger than $100$. One should note that the proposed CG type algorithms have a much lower computational complexity than LSMI, AVF and MWF algorithms.

\begin{figure}[!htb]
\centering
\includegraphics[width=86mm]{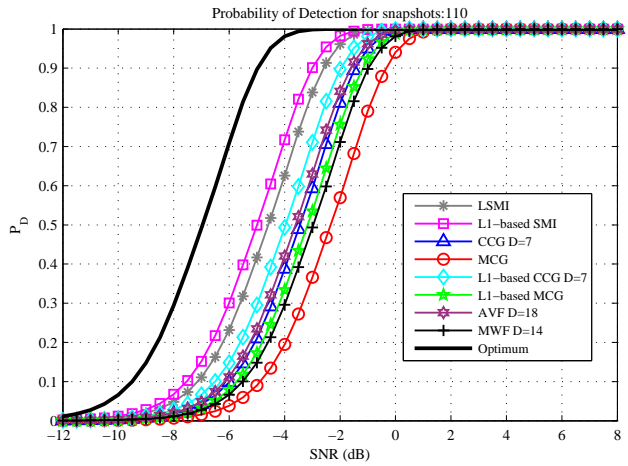}
\caption{
Probability of detection performance versus SNR with $K=110$ snapshots. $P_{fa}=10^{-6}$ and the other parameters are the same as the second example.} \label{pd_snr_homo}
\end{figure}

In the third example, we present the probability of detection $P_d$ versus SNR with the target injected at the azimuth of $0^\circ$ and Doppler frequency $100$Hz in Fig.\ref{pd_snr_homo}. We assume the false alarm rate $P_{fa}$ is set to $10^{-6}$ and the number of the secondary data is $K=110$. The plots illustrate a similar trend to the SINR loss performance in the second example. Note that we obtain a performance gain of about $1$dB in terms of SNR for $l_1$-based CG type algorithms, as compared with conventional CG type algorithms.

\begin{figure}[!htb]
\centering
\includegraphics[width=86mm]{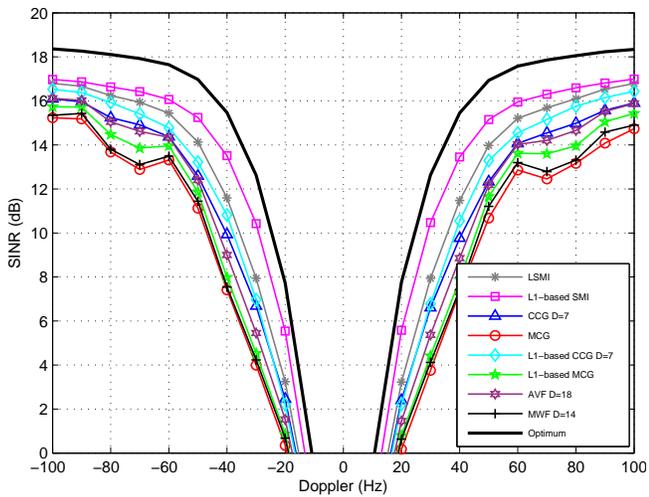}
\caption{
SINR performance against Doppler frequency with $K=160$ snapshots and Doppler frequency space from $-100$ to $100$Hz. The other parameters are the same as the second example.} \label{sinr_doppler_homo}
\end{figure}

Fig.\ref{sinr_doppler_homo} shows the SINR performance against the target Doppler frequency at the azimuth of $0^\circ$ with a total of $K=160$ snapshots. Here, we suppose the potential Doppler frequency space is from $-100$Hz to $100$Hz. The parameters of all algorithms are the same as the second example. The curves in the figure demonstrate a similar trend to the results of previous examples. Additionally, the $l_1$-based SMI algorithm displays much better performance to the slow targets than other algorithms.

\subsection{Measured Data}
In this section, we apply the proposed algorithms to the Mountain-Top data set. This data was collected from commanding sites (mountain tops) and radar motion is emulated using a technique developed at Lincoln Laboratories \cite{Titi1996,Peckham2000}. The sensor consists of $14$ elements and the data are organized in CPIs of $16$ pulses. Here, we use the data file $t38pre01v1$ CPI$6$, which could be obtained from the internet \cite{mountiantop}. The pulse PRF was $625$Hz and the instance bandwidth after pulse compression was $500$kHz. There are $403$ independent range samples available for the training data support. The clutter was located around $245^\circ$ azimuth and the target was at $275^\circ$ with a Doppler frequency $156$Hz. All the data processed following are through pulse compression firstly. Note that the clutter and target have the same Doppler frequency, hence separation is impossible in the Doppler domain but possible in the spatial domain. The estimated angle-Doppler profile using all $403$ samples is given in Fig.\ref{spectrum_mountaintop}, which shows a serious heterogeneity.

Fig. and Fig. display the STAP output power of all algorithms in the
range of $147$-$162$ km. Here, the interference covariance matrix
estimated using a symmetric sliding window with a total of $20$
snapshots for Fig. and $40$ snapshots for Fig.. For each CUT, the
snapshots do not include the $6$ snapshots around the CUT. In the
figures, we also give the unadapted weight vector, which equals the
steering vector ${\bf w}={\bf s}$. We see that the target is clearly
not detectable without adaptive processing. To have a clear
comparison amongst different algorithms, we show the differences
between the output power at $154$ km and the next highest power peak
in Table \ref{mountainoutput}, where ''-'' presents the target not
detectable. Here, $6$ range bins around the rang bin of the target
is not used for comparison since they are the guide cells. Seen from
the table, we find that: (1) the proposed $l_1$-based SMI algorithm
obtains the best detection performance in both situations, which is
the same conclusion as that using simulated data; (2) the proposed
$l_1$-based CG type algorithms obtain better performance than the
conventional CG type algorithms (although the proposed $l_1$-based
MCG algorithm has a pseudo target at the range $153$km, when the
secondary data record is $20$ snapshots, the conventional MCG
algorithm can not detect the target at all.); (3) the proposed
$l_1$-based CCG algorithm outperforms AVF and MWF algorithms in both
situations. Hence, we can conclude that our proposed algorithms show
a robust performance in heterogeneous environments.

\section{Conclusions}
In this paper, we have proposed novel SA-STAP algorithms with $l_1$-norm regularization for targets detection in airborne radar systems. The proposed SA-STAP algorithms employed a sparse regularization to the MV cost function to exploit the sparsity of the received data and filter weights. To solve this kind of optimization problem, an $l_1$-based SMI algorithm was directly developed, but it required matrix inversion resulting in a high computational cost. Accordingly, we have proposed low-complexity SA-STAP algorithms based on CG techniques. A detailed analysis of the computational complexity and the performance of the SA-STAP algorithms were carried out. Simulation results with both simulated and measured data showed that the proposed algorithms outperformed conventional algorithms and exhibited a robust performance in heterogeneous environments.

\appendices


\begin{table}[ht]
  \centering
  \caption{The $l_1$-based CCG Algorithm}\label{tabel.ccg}
  \small
  \begin{tabular}{l}
  \hline
  \textbf{Initialization:}\\
  $\hat{{\bf R}[0]} = \delta{\bf I}$,
  ${\bf v}[0] = {\bf s}$,$\eta_{\rm CCG}$,\\
  \hline
  \textbf{Recursion: For each snapshot $k=1,\cdots,L$}\\
  \hspace{1em}\textbf{STEP 1: Start:}\\
  $\hat{{\bf R}}[k] = \beta\hat{{\bf R}}[k-1] + {\bf x}[k]{\bf x}^H[k]$,\\
  ${\boldsymbol \Lambda}[k] = \mbox{diag}\left\{\frac{1}{\left|w_1[k-1]\right|+\epsilon},\cdots,\frac{1}{\left|w_{NM}[k-1]\right|+\epsilon}\right\}$,\\
  ${\bf G}[k] = \hat{{\bf R}}[k] + \lambda{\boldsymbol \Lambda}[k]$,\\
  ${\bf g}_0[k] = {\bf s} - {\bf G}[k]{\bf v}_0[k]$,
  ${\bf p}_1[k] = {\bf g}_0[k]$,
  $\rho_0[k] = {\bf g}^H_0[k]{\bf g}_0[k]$,\\
  \hspace{1em}\textbf{STEP 2: For $d=1,\cdots,D$ and $\rho_{d-1}[k]>\eta_{\rm CCG}$}\\
  ${\bf z}_d[k] = {\bf G}[k]{\bf p}_d[k]$,\\
  $\alpha_d[k] = \left[{\bf p}_d^H[k]{\bf z}_d[k]\right]^{-1}\rho_{k-1}[k]$,\\
  ${\bf v}_d[k] = {\bf v}_{d-1}[k-1]+\alpha_d[k]{\bf p}_d[k]$,\\
  ${\bf g}_d[k] = {\bf g}_{d-1}[k]-\alpha_d[k]{\bf z}_d[k]$,\\
  $\rho_d[k] = {\bf g}^H_d[k]{\bf g}_d[k]$,\\
  $\nu_d[k] = \frac{\rho_d[k]}{\rho_{d-1}[k]}$,\\
  ${\bf p}_{d+1}[k] = {\bf g}_d[i] + \nu_d[k]{\bf p}_d[k]$,\\
  \hspace{1em}\textbf{STEP 3: After end STEP 2}\\
  ${\bf v}_0[k+1] = {\bf v}_d[k]$,\\
  ${\bf w}[k] = \frac{{\bf v}_0[k]}{{\bf s}^H{\bf v}_0[k]}$,\\
  \hline
  \textbf{Final output:}\\
  $y[k] = {\bf w}^H[k]{\bf x}[k]$.\\
  \hline
   \end{tabular}
\end{table}

\begin{table}[ht]
  \centering
  \caption{The $l_1$-based MCG Algorithm}\label{tabel.mcg}
  \begin{small}
  \begin{tabular}{l}
  \hline
  \textbf{Initialization:}\\
  $\hat{{\bf R}[0]} = \delta{\bf I}$,
  ${\bf v}[0] = {\bf s}$,
  ${\bf w}[0] = {\bf s}$,
  ${\bf g}[0] = {\bf s}$,
  ${\bf p}[1] = {\bf s}$,\\
  \hline
  \textbf{Recursion: For each snapshot $k=1,\cdots,L$}\\
  $\hat{{\bf R}}[k] = \beta\hat{{\bf R}}[k-1] + {\bf x}[k]{\bf x}^H[k]$,\\
  ${\boldsymbol \Lambda}[k] = \mbox{diag}\left\{\frac{1}{\left|w_1[k-1]\right|+\epsilon},\cdots,\frac{1}{\left|w_{NM}[k-1]\right|+\epsilon}\right\}$,\\
  ${\bf G}[k] = \hat{{\bf R}}[k] + \lambda{\boldsymbol \Lambda}[k]$,\\
  $\alpha[k] = \left[{\bf p}^H[k]{\bf G}[i]{\bf p}[k]\right]^{-1}$\\
  \qquad\qquad$\times\left\{\beta \left[{\bf p}^H[k]{\bf g}[k-1]-{\bf p}^H[k]{\bf s}\right] - \mu{\bf p}^H[k]{\bf g}[k-1]\right\}$,\\
  ${\bf v}[k] = {\bf v}[i-1]+\alpha[i]{\bf p}[i]$,\\
  ${\bf g}[k] = \left(1-\beta\right){\bf s} + \beta{\bf g}[k-1] - \alpha[k]{\bf G}[k]{\bf p}[k]$\\
  \qquad\qquad$-\left\{\lambda\left[1 - \beta\right]{\boldsymbol \Lambda}[k] + {\bf x}[k]{\bf x}^H[k]\right\}{\bf v}[k-1]$,\\
  $\nu[k] = \frac{\left[{\bf g}[k]-{\bf g}[k-1]\right]^H{\bf g}[k]}{{\bf g}^H[k-1]{\bf g}[k-1]}$,\\
  ${\bf p}[k+1] = {\bf g}[k-1] + \nu[k]{\bf p}[k]$,\\
  ${\bf w}[k] = \frac{{\bf v}[k]}{{\bf s}^H{\bf v}[k]}$,\\
  \hline
  \textbf{Output:}\\
  $y[k] = {\bf w}^H[k]{\bf x}[k]$.\\
  \hline
   \end{tabular}
   \end{small}
\end{table}

\begin{table}[ht]
  \centering
  \caption{Comparison of the Computational Complexity}\label{table.complexity}
  \begin{small}
  \begin{tabular}{|c|c|c|}
  \hline
  Algorithm & Additions & Multiplications \\
  \hline
   LSMI & $O\left(L^3\right)+O\left(L^2\right)$ & $O\left(L^3\right)+O\left(L^2\right)$\\
  \hline
   $l_1$-based SMI & $O\left(L^3\right)+O\left(L^2\right)$ & $O\left(L^3\right)+O\left(L^2\right)$ \\
  \hline
   MWF & $ DL^2+(4D-D^2)L$ & $2DL^2+(5D-D^2)L$\\
   & $+\frac{D^3}{3}-\frac{3D^2}{2}-\frac{D}{3}$ & $+\frac{2D^3}{3}-2D^2+\frac{16D}{3}$\\
  \hline
   AVF & $ (2D+1)L^2+(4D$ & $2(D+1)L^2 $\\
       & $+1)L-4D-1$ & $+7DL+L$\\
  \hline
   CCG & $ (D+2)L^2+(4D$ & $(D+3)L^2$\\
       & $+2)L-2D-2$ & $+5DL+3L$\\
  \hline
   MCG & $3L^2+10L-4$ & $4L^2+13L+2$\\
  \hline
   $l_1$-based CCG& $(D+3)L^2+(4D+$ & $(D+3)L^2$\\
    & $3)L-2D-2$ & $+7DL+L$ \\
  \hline
   $l_1$-based MCG& $5L^2+11L-4$ & $4L^2+13L+2$\\
  \hline
   \end{tabular}
   \end{small}
\end{table}

\begin{table}[ht]
  \centering
  \caption{Results of Mountain Top data}\label{mountainoutput}
  \begin{small}
  \begin{tabular}{|c|c|c|}
  \hline
  Algorithms & $20$ snapshots & $40$ snapshots \\
  \hline
   unadapted &  - & -  \\
  \hline
    LSMI &  $7.1$dB &  $12.4$dB \\
  \hline
    $l_1$-based SMI & $9.3$dB &  $15.1$dB \\
  \hline
    MWF & $4.6$dB &  $10.0$dB \\
  \hline
    AVF & $5.6$dB &  $11.9$dB \\
  \hline
    CCG & $5.2$dB & $11.9$dB  \\
  \hline
    MCG & - &  $6.2$dB \\
  \hline
    $l_1$-based CCG & $7.1$dB & $13.0$dB  \\
  \hline
    $l_1$-based MCG & $0$dB & $9.5$dB  \\
  \hline
   \end{tabular}
   \end{small}
\end{table}


\end{document}